# A topological perspective into the sequence and conformational space of proteins

K.Silpaja Chandrasekar and M.V Sangaranarayanan*
Department of Chemistry
Indian Institute of Technology-Madras Chenni-600036 India

Abstract

The precise sequence of aminoacids plays a central role in the tertiary structure of proteins and their functional properties. The Hydrophobic-Polar lattice models have provided valuable insights regarding the energy landscape. We demonstrate here the isomorphism between the protein sequences and designable structures for two and three dimensional lattice proteins of very long aminoacid chains using exact enumerations and intuitive considerations. It is customarily thought that computations of this magnitude will require 'million years' with the present day computers. We emphasize that the topological arrangement of the aminoacid residues alone is adequate to deduce the designable and non-designable sequences without explicit recourse to energetics and degeneracies. The results indicate the computational feasibility of realistic lattice models for proteins in two and three dimensions and imply that the fundamental principle underlying the designing of structures is the connectivity of the hydrophobic and polar residues.

*E-mail: sangara@iitm.ac.in

# 1. Introduction

The prediction of tertiary structures of proteins from the constituent aminoacid sequences constitutes a major challenging problem in science, on account of its complexity and importance[1].Despite considerable efforts in unraveling the mysterious connection between the native structures of proteins and their most stable sequences, the mechanism underlying the energy landscape is not yet clear[2].However, valuable insights have hitherto been gathered from minimalistic lattice models of proteins and simple fundamental rules seem to govern the native structures. The highly popular Hydrophobic-Polar (HP) model[3] for a chain of N aminoacid residues (or nodes) has provided information regarding the designable protein sequences *vis a vis* conformations for two and three dimensional lattices of moderate chain lengths. The lattice models provide a basis for comprehending the ground state of the native structures of proteins if systematically and exhaustively analyzed. The exact enumeration has been recognized as an NP hard problem[4] which restricts the conformational analysis to chains of small lengths. Further, the tertiary structures of proteins is dictated solely by the HP sequence [5]and a recent *experimental* evidence highlights the importance of the sequence alignment using a protein-like polymer sequence for the coil to globule transition[6]. It is therefore imperative to carry out an *exact* sequence-conformation mapping from a new perspective for fairly long aminoacid chain lengths. The applicability of spin glass models to coil to globule transitions has also been demonstrated[7]. In the case of protein sequences, the *energies and degeneracies* of $2^N$ states need to be computed for identifying the native states. This can be carried out either using Monte Carlo simulations[8] or graph theoretical procedures[9].

Here, we report a new method of comprehending the conformational and sequence spaces of proteins for typical two and three dimensional lattices within a *modified* HP model framework, the chain lengths considered here for exact analysis being noticeably very large. In contrast to the hitherto-known lattice models which emphasize the pairwise interaction energies of various contacts, the present analysis demonstrates that the mere enumeration of the topological arrangements is adequate to decipher the native sequences and conformations. While the number of hydrophobic-hydrophilic contacts are exactly enumerated here for two-dimensional square lattices (with aminoacid chain lengths N = 64,144 and 256) and cubic lattices (with N =27 and 64), the *most designable and non-designable sequences* are deduced for N = 16; N =25 and N= 36 in two dimensions and for N =27 and N= 64 in three dimensions.

**Enumeration of topological arrangement of protein sequences**

The enumeration of the black-white edges {A(p,q)} for large lattice sizes in two and three dimensions is a challenging task and plays a central role in the calculation of the associated partition functions[10]. When this counting is accomplished, the number of ways of obtaining different hydrophobic-hydrophilic contacts becomes available for all compositions of proteins. Hence a high level of discrimination among the protein sequences arises in this approach due to the exact counting of the hydrophobic-hydrophilic contacts. As is well-known, different compositions lead to dissimilar sequences[11]. Furthermore, the exact analysis pertaining to long chains is essential in order to investigate the cooperative phenomena[12].

**(A) Square lattice**

Elsewhere, the counting of the black-white edges for the estimation of the partition function of two-dimensional nearest neighbor Ising models has been reported in the case of a square lattice of 36 sites[13]. We have carried out here this enumeration for the chain lengths (N= 64; 144 and 256) for each composition of the protein using the *complete sequence space* ($2^N$) and Fig 1 depicts the dependence of A(p,q) on the fraction of the hydrophobic residues (p/N) and H-P contacts (q/N) for a square lattice of 64 sites, the numbers being ferreted out from the exact enumeration. There is a symmetry in the entries of A (p, q) at the midpoint viz p=N/2 indicating that the labeling of the H and P residues is interchangeable. The total number of the arrangements ($\sum$ A(p,q)) is maximum when the ratio of hydrophobic to polar residues is unity(i.e. p = N/2) whatever be the chain length and it is well known that unique optimum embedding occurs when the H/P ratio is nearly unity[14]. When N is odd, the frequency of occurrence is maximum at p=(N-1)/2.It is inferred that for a square lattice of 16 aminoacid chains, when the number of hydrophobic residues (p)is 12, five different arrangements are possible with q = 8;10;12;14 and 16 , consistent with an earlier analysis[15] .

**(B)Simple cubic lattice**

While the methodology for simple cubic lattices is identical as for square lattices, the maximum value of N was fixed as 64 and Fig 1b depicts the variation of A (p, q) with the fraction of the hydrophobic residues (p/N) and H-P contacts (q/N) for a simple cubic lattice of 64 sites, as an illustration.

**Identification of favorable protein sequences**

In the original HP model[3], the origin of the native structures consists in the favorable negative interaction energy between H-H contacts ($E_{H-H}$) while the H-P ($E_{H-P}$) and P-P ($E_{P-P}$) energies were assigned zero. Subsequently, other parametrization schemes for the interaction energies were envisaged[16-18].An alternate method advocated in dimensionless units[19] is the following: $E_{H-H}$ = -2.3; $E_{H-P}$ = -1 and $E_{P-P}$= 0.Thus, the elucidation of the precise energetic factors which control the design of protein sequences is a non-trivial endeavor.

**(A)Square lattice**

For any fraction of the hydrophobic residues (p/N), the arrangements with the least number of H-P contacts can be deciphered. This number represents the upper limit of the H-P contacts for any favorable sequence. If the value of p is chosen as N/2 for brevity, q ≥8 (for N=16), q ≥ 12 (for N=36), q≥ 16(for N=64), q≥20 (for N =144) and q ≥ 24 (for N=256) are considered as non-*favorable*. This assumption is consistent with the fact that an alternating sequence of polar and non-polar aminoacids hinders protein folding [20].Typically, in the case of 16-residue aminoacids, the total number of sequences is 12870 when p=8 and hence the percentage of favorable sequences is 0.062.While the total number of sequences (viz 12,870) can easily be obtained using the binomial theorem, the precise manner in which these 12,870 sequences comprise different types of black-white edges is required for identifying the favorable and non-favorable sequences. The identification of such favorable sequences is feasible for all compositions of proteins till N= 256 for square lattices.

**(B) Simple cubic lattice**

As in the above square lattice case, the arrangements wherein $q \geq 14$ (for $N=3^3$) and $q \geq 18$ (for $N=4^3$), are considered *non-favorable sequences* for respective p/N values of 13/27 and 32/64. We emphasize that the enumeration of the hydrophobic-polar contacts is essentially a preliminary step, since the designable sequences should take into account, (i) the self-avoiding nature; (ii) the hydrophobic-polar ratio and (iii) the connectivity of the favorable sequence(s). This *inter alia* requires the enumeration of the conformations in two and three dimensions, confined to the lattice.

**Enumeration of all conformations**

The conformations of lattice proteins are often analyzed using two or three dimensional self-avoiding random walks[21] and these estimates are essential to deduce the percentage of designable conformations. The composition of the lattice protein is given by p while q is a measure of the topological contact; hence the designable sequences need to be encoded, consistent with the favorable sequence(s).

**(A) Square lattice**

We have enumerated the total number of conformations for aminoacid residues ranging from N =4 to N = 51(Table 1). While the conformations for N till 25 have been enumerated earlier[22], those for N > 25, have been provided here for the first time. However, the scaling arguments using critical exponents can yield estimates till N =71 by extrapolation[23].

**(B) Simple cubic lattice**

We have enumerated the number of conformations for N ranging from 4 to 40 in the case of simple cubic lattices (Table 2). The hitherto-known estimates [24] till N = 36 have been reproduced apart from new exact values till N=40 here.

**Enumeration of the Designable and Non-designable Sequences**

The foregoing analysis indicates a method of identifying favorable and non-favorable sequences for square and simple cubic lattices, employing the information on H-P contacts. A given structure can in principle be designed from a large number of *designing sequences* ($S_N$) and $D_N$ denotes the number of *designable conformations* for chains of length N.

**(A) Square lattice**

We establish here (Methods Section), the precise isomorphism between the favorable sequence (identified from the least number of black-white edges $q_{min}$) and the designing sequence (enumerated from the constraint { $q \leq (q_{min} -1)$}). These values computed here (for p/N =0.5) solely from the topological contacts, are entirely in agreement with the hitherto-known values till N = 25 ) while $S_N$ for N > 25 are provided here for the first time(Table 3). Hence it follows that all the hitherto-known values of $S_N$ pertain to the H/P ratio of unity with the earlier-mentioned constraint on the topological contacts {$q \leq (q_{min} -1)$} being implicitly present and it is of interest to enquire the influence of H/P ratio on these. As anticipated, the maximum value of $S_N$ occurs when H/P =1. From $S_N$, the *most designable sequences* (Table 4) are further deciphered by invoking the 'adjacent sites' assumption (Methods Section) and the flowchart shown in Fig 2 illustrates the *modus operandi* involved in the elucidation for N =36.The other structures are designated as less designable. We note that the optimal structure for N =36 earlier deduced using a deterministic optimization approach[25] and a systematic designability criterion within a modified HP model[26] is consistent with the 'adjacent sites' assumption invoked here. A combined approach using genetic algorithm in conjunction with Taguchi method has also been studied[27] for N=64. Figs 3, 4 and 5 depict respectively the most designable sequences for N=16, N=25 and N=36. Figs 6, 7 and 8 depict respectively the non - designable sequences for N=16, N=25 and N=36.

**(B)Simple cubic lattice**

The *non-favorable* sequences are those having H-P contacts $\geq q_{min}$ for any chosen hydrophobic ratio. In view of the lattice coordination number being six for simple cubic lattices, the earlier *ansatz* of estimating the self-avoiding walks with $q \leq (q_{min} -1)$ is not valid for obtaining the designing sequences. Since protein folding is predominantly associated with the hydrophobic phenomena it is appropriate to search for the least number of H-P contacts whereby the natural choices for $q_{min}$ are zero and unity for any composition of the protein. These values computed using self-avoiding walks under the constraint $q \leq 1$ (for p/N =0.5) enumerated here (Table 5) solely from the H-P contacts, are entirely in agreement[28] with the hitherto-available values (till N=19). However, $S_N$ for *larger* aminoacid chain lengths (N >19) have been reported here for the first time. The total number of designable sequences for 4*4*4 lattice is 389 and the flow chart in Fig 2 depicts the methodology for deducing the most designable sequences. A Monte Carlo simulation incorporating the solvent-aminoacid interaction energies has been investigated for calculating the folding time and melting temperatures[29] restricting the analysis to 27 aminoacid chains. In a hybrid method involving genetic algorithm and particle swarm optimization, the local search method with six internal coordinates has been invoked[30]. The influence of H/P ratio on $S_N$ can also be deciphered as before and its value is maximum when the fraction of the hydrophobic residues is 1/2. Figs 9 and 10 depict respectively the most designable sequences for N=27 and N=64. Table 6 provides the partitioning of $S_N$ into the most designable and less designable sequences for N ranging from 4 to 27 and for N=64.Figs 11 and 12 illustrate respectively typical non - designable sequences for N=27 and N=64.

The foregoing analysis indicates that the precise counting of the topological contacts between the hydrophobic and polar residues for any lattice proteins is adequate to decipher their designable and non-designable sequences in two and three dimensions. Thus, the connectivity dictates the most designable protein structure as has been noted earlier[31]. The number of designable sequences is intimately linked to the composition of the proteins. Hence the hitherto-limited success in the protein folding problem may be not due to the limitation associated with the exhaustive searching of the sequence and conformational energy space.

While the final test regarding the accuracy of predicting protein structures from the constituent aminoacid sequences requires the analysis of the actual database[32], the method propounded here indicates the computational feasibility of lattice proteins as well as the existence of a general robust practical method for exactly enumerating the designable and non-designable sequences.

**Methods**

We assume that a given protein consists of various compositions of hydrophobic (H) and polar (P) residues, spanning a lattice of N sites as in the original H-P model [3]; however, no pair-wise interaction energies have been explicitly invoked here. We consider here two-dimensional square and three-dimensional simple cubic lattices. Either a hydrophobic (designated as 'black') or hydrophilic part (denoted as 'white') occupies each vertex. The symbols 'p' and 'q' denote respectively the number of the hydrophobic residues and hydrophobic-hydrophilic contacts (black-white edges). The combinatorial problem consists in counting all the arrangements of the black-white edges viz {A (p, q)}. In order to facilitate extrapolation to infinite sites, periodic boundary conditions are assumed as is customary in statistical physics[33], Monte Carlo simulation[34] and protein folding[35]. This enumeration yields all the hydrophobic-hydrophilic contacts {A (p, q)} when the fraction of the hydrophobic residues (p/N) varies from 0 to 1. We have adopted a multi-step strategy to deduce the interdependence of sequences and conformations of proteins. Firstly, we enumerate exactly all the hydrophobic-hydrophilic contacts in lattice proteins of various compositions and this yields the topological alignment of sequences and their statistical weights. Secondly, we identify the most favorable sequence *vis a vis* native state of proteins for any given hydrophobic to polar ratio by identifying the number of arrangements with the least number of black-white edges ($q_{min}$). The sequences wherein $q \geq q_{min}$ are considered *non-favorable* for a chosen p/N value. Thirdly, for this *composition,* we compute the number of designing sequences ($S_N$) by counting all the self-avoiding walks for the H-P contacts when (A) $q \leq q_{min} -1$ for square and (B) $q \leq 1$ for simple cubic lattices. Fourthly, we infer the most designable and less designable sequences for square lattices by postulating that the starting and destination sites should be either on the same row or column ('adjacent' arrangements) This 'adjacent' arrangement has the highest frequency of occurrence. The identification of the most designable sequences in simple cubic lattices is more involved due to the coordination number of the lattice being six; hence we follow a different approach wherein six different structures designated as $p_1$, p2, p3, p4, p5 and $p_6$ have been invoked.

**Acknowledgements**

The financial support by the Department of Science and Technology, Government of India is gratefully acknowledged.

**Fig 1**: Exact enumeration of the number of ways of obtaining hydrophobic - hydrophilic contacts {A (p, q)} for a chain length of 64 aminoacids for any composition of lattice protein: (a) square and (b) simple cubic lattice. Typically for (p/N) = 0.0625, in (a), the fraction of hydrophobic – hydrophilic contacts (q/N) is 0.125, 0.156, 0.187, 0.218 and 0.25; the corresponding total number of ways of obtaining these are 64; 1152; 575408; 33920 and 24832. In (b), the fraction of hydrophobic – hydrophilic contacts (q/N) is 0.218, 0.281, 0.343 and 0.406; the total number of ways of obtaining these are respectively 324; 4096; 4056 and 248064 respectively. The complete set of $2^{64}$ values for all compositions has been enumerated.

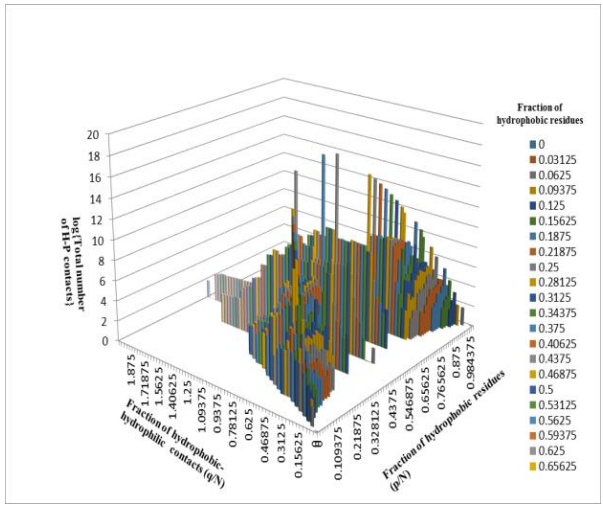

(a)

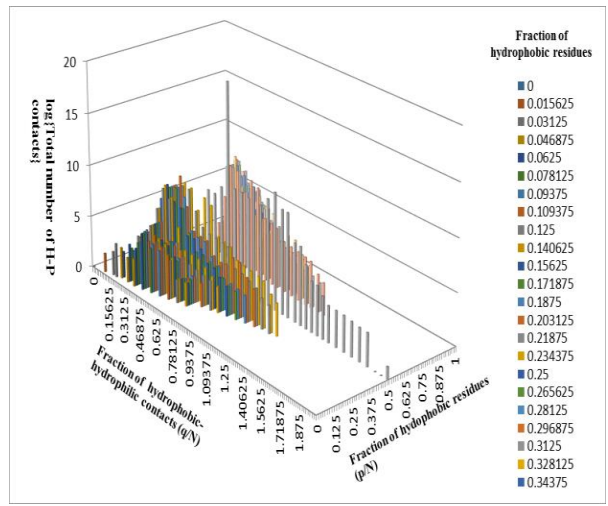

(b)

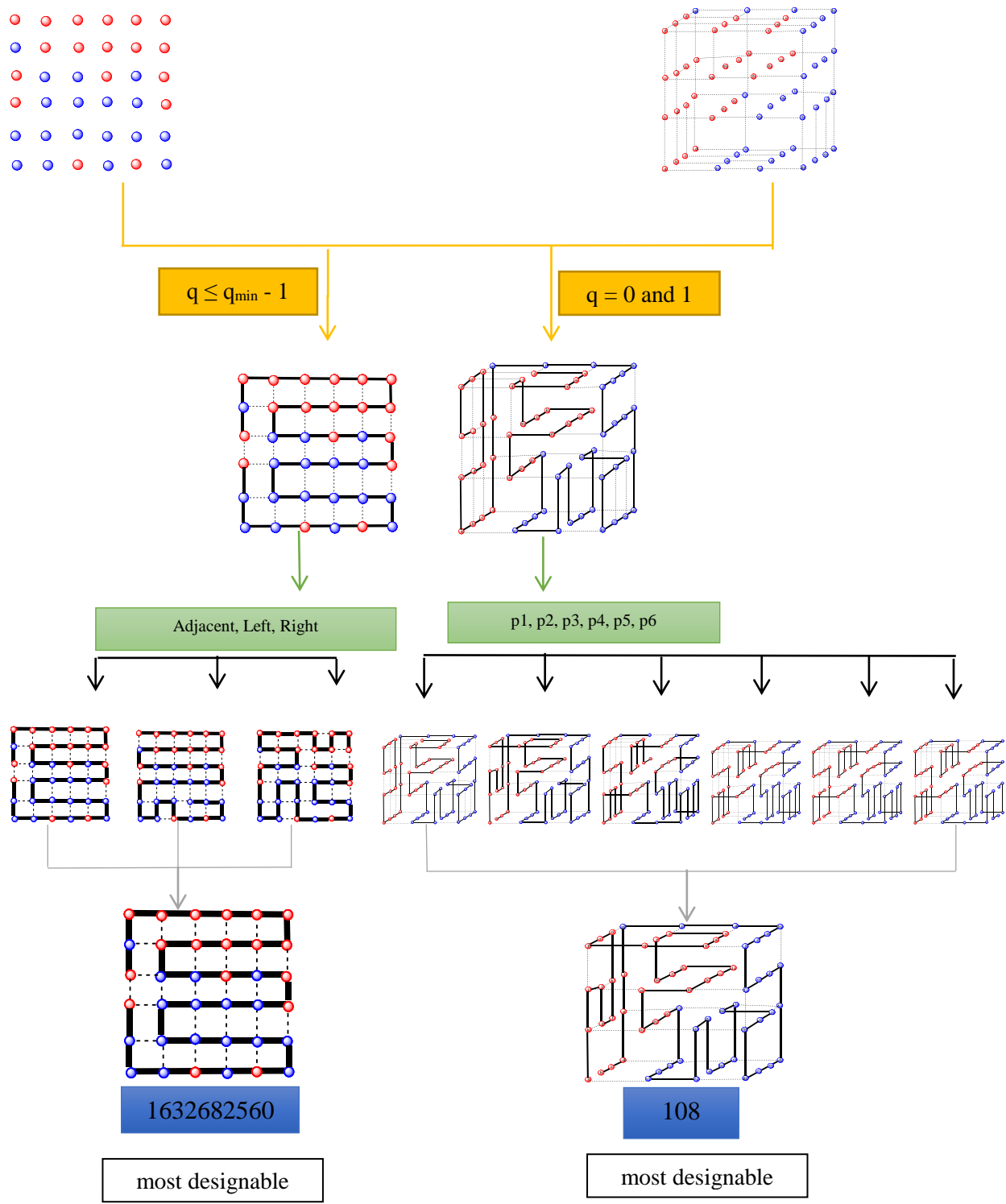

**Fig 2**: Elucidation of the most designable sequences for a chain length of N aminoacids with equal number of hydrophobic and hydrophilic residues: (a) square lattice with N =36 and (b) simple cubic lattice with N =64.

**Fig 3:** Different designable sequences of a two-dimensional square lattice of 16 amino acids with equal number of hydrophobic and hydrophilic residues: (a) most designable, (b) and (c) less designable sequences. The frequency of occurrence of a, b and c are 1027, 256 and 256 respectively. The blue and red circles denote respectively the hydrophobic and hydrophilic parts.

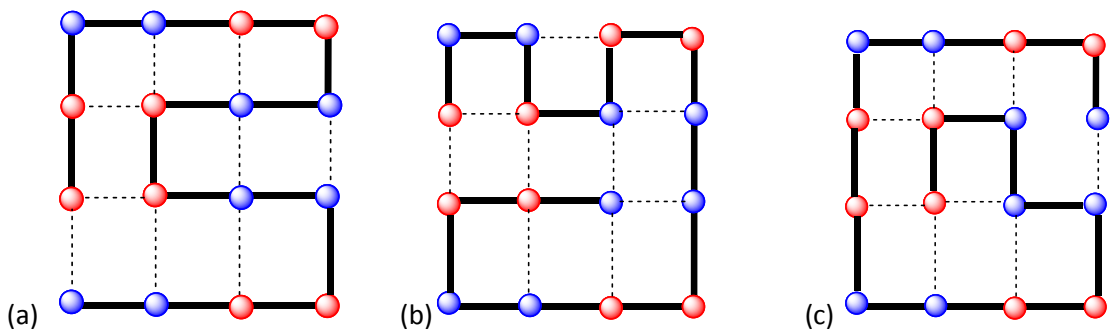

**Fig 4:** Different designable sequences of a two-dimensional square lattice of 25 amino acids with equal number of hydrophobic and hydrophilic residues: (a) most designable, (b) and (c) less designable sequences. The frequency of occurrence of a, b and c are 236973, 264087 and 264087 respectively. The blue and red circles denote respectively the hydrophobic and hydrophilic parts.

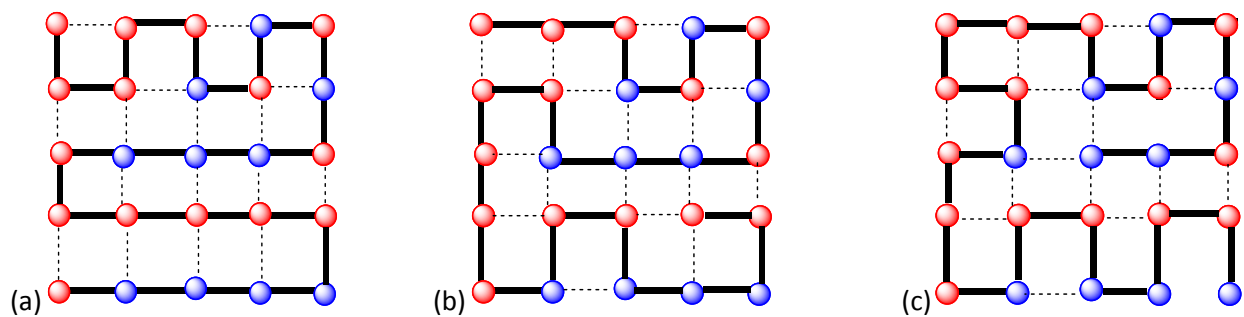

**Fig 5:** Different designable sequences of a two-dimensional square lattice of 36 amino acids with equal number of hydrophobic and hydrophilic residues: (a) most designable, (b) and (c) less designable sequences. The frequency of occurrence of a, b and c are 1632682560, 60466176 and 60466176 respectively. The blue and red circles denote respectively the hydrophobic and hydrophilic parts.

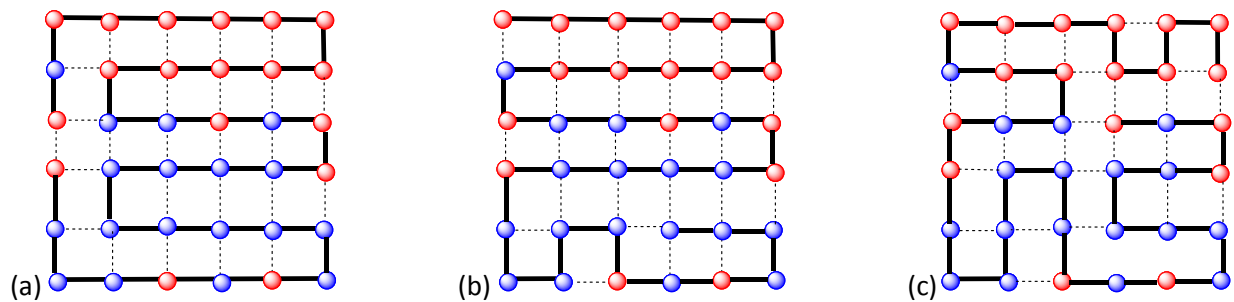

**Fig 6:** Typical non – designable sequences for a two – dimensional square lattice of 16 aminoacids with equal number of hydrophobic and hydrophilic residues. The total number of non - designable sequences is 241 wherein $q \geq q_{min}$

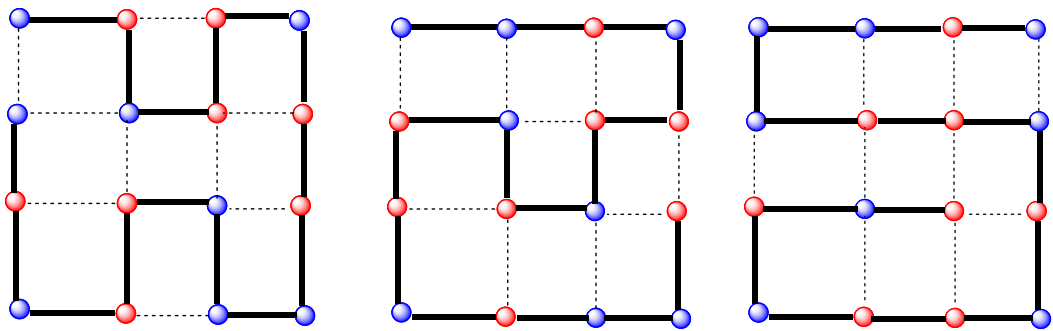

**Fig 7:** Typical non – designable sequences for a two – dimensional square lattice of 25 aminoacids with equal number of hydrophobic and hydrophilic residues. The total number of non - designable sequences is 577, wherein q ≥q $_{min}$

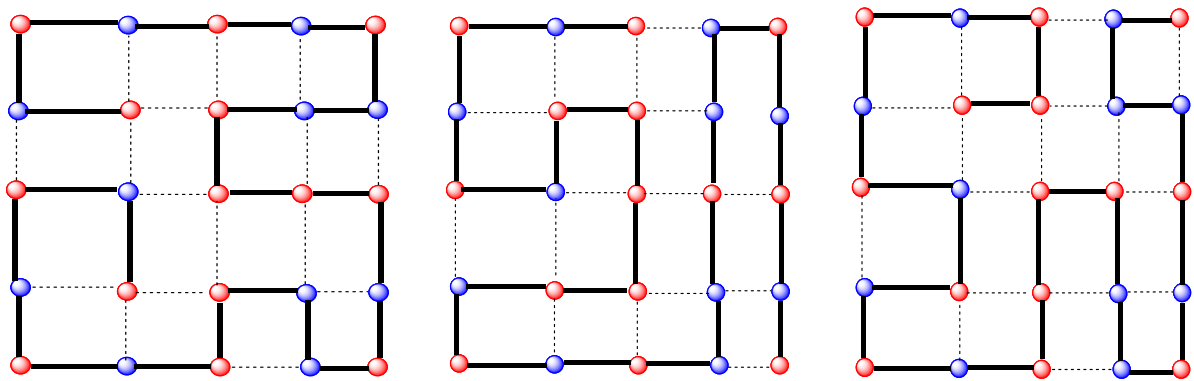

**Fig 8:** Typical non – designable sequences for a two – dimensional square lattice of 36 aminoacids with equal number of hydrophobic and hydrophilic residues. The total number of non - designable sequences is 21655, wherein q ≥ q $_{min}$

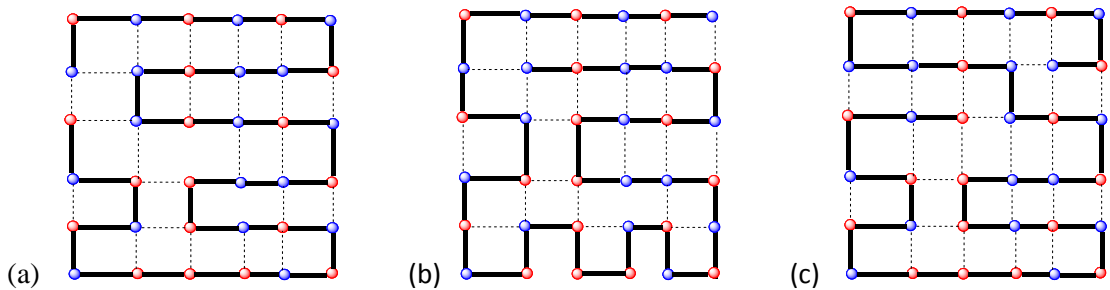

**Fig 9**: Different designable sequences of a simple cubic lattice of 27 amino acid chains with equal number of hydrophobic and hydrophilic residues. p4 denotes a typical most - designable sequence while p1, p2, p3 and p5 denote less designable sequences. The frequency of occurrence of p1, p2, p3, p4 and p5 are 22, 35, 16, 60 and 24.

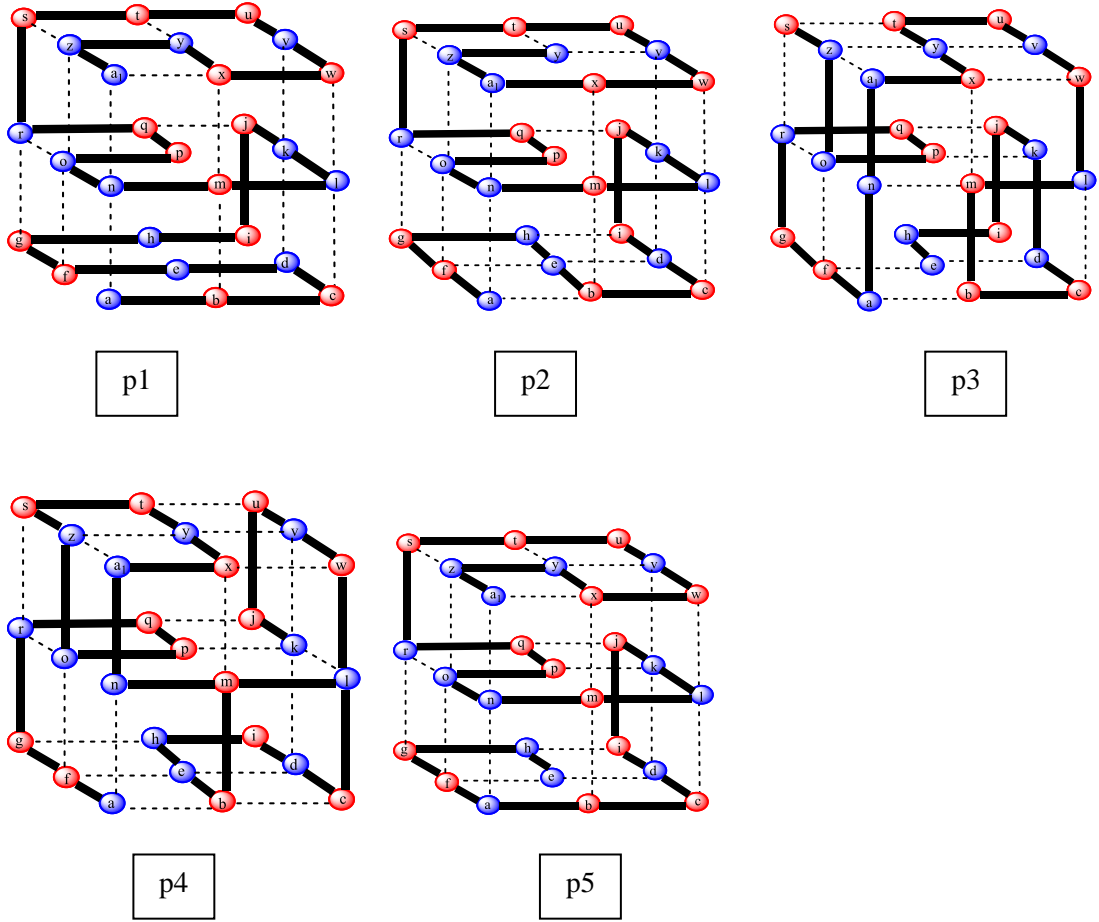

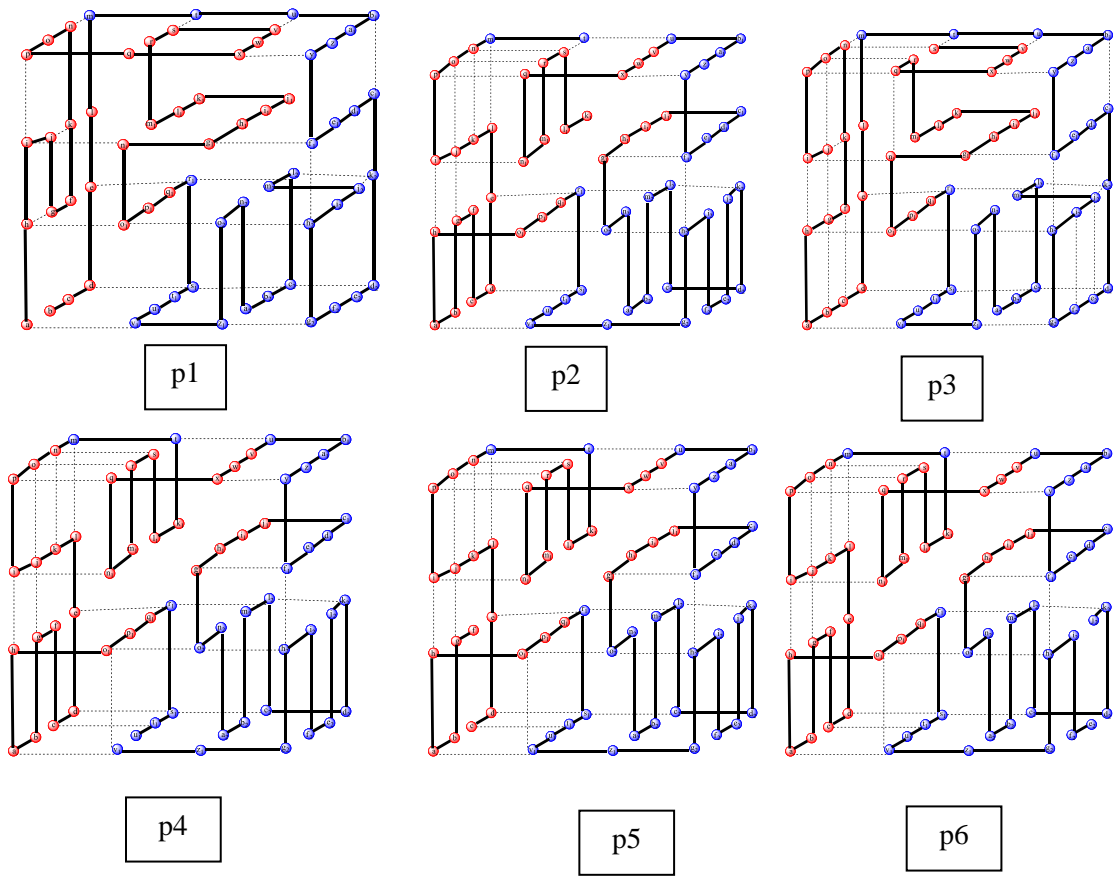

**Fig 10**: Different designable sequences of a simple cubic lattice of 64 amino acid chains with equal number of hydrophobic and hydrophillic residues. p2 denotes a most - designable sequence while p1, p3, p4, p5 and p6 denote less designable sequences. The frequency of occurrence of p1, p2, p3, p4, p5 and p6 are 64, 108, 8, 47, 4 and 50.

**Fig 11:** Typical non–designable sequences for a simple cubic lattice of 27 amino-acids with equal number of hydrophobic and hydrophilic residues. The total number of non - designable sequences is 121 wherein q >1.

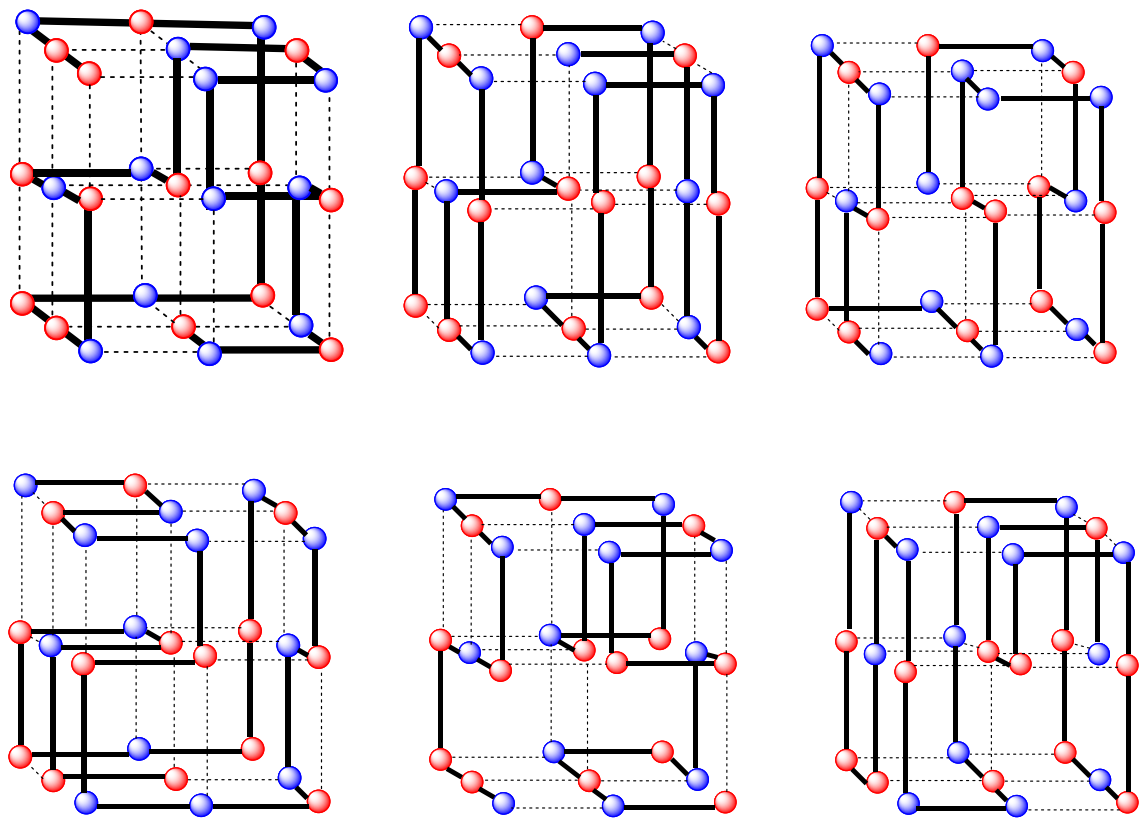

**Fig 12:** Typical non – designable sequences for a simple cubic lattice of 64 amino-acids with equal number of hydrophobic and hydrophilic residues. The total number of non - designable sequences is 173 wherein q>1.

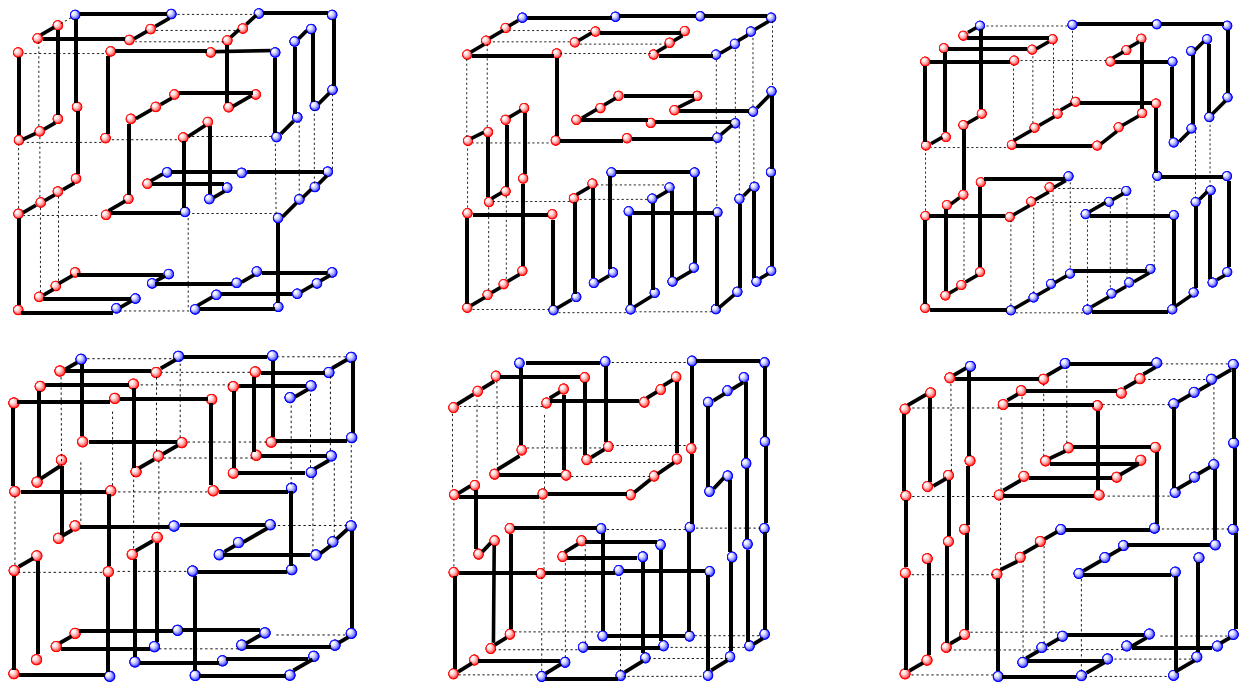

*Table 1*- Total number of conformations in square lattices for various chain lengths (N).

| Number of Amino acid residues (N) | Total Number of conformations |
|---|---|
| 4 | 5 |
| 5 | 13 |
| 6 | 36 |
| 7 | 98 |
| 8 | 272 |
| 9 | 704 |
| 10 | 2034 |
| 11 | 5513 |
| 12 | 15037 |
| 13 | 40617 |
| 14 | 110188 |
| 15 | 296806 |
| 16 | 802075 |
| 17 | 2155667 |
| 18 | 5808335 |
| 19 | 15582342 |
| 20 | 41889578 |
| 21 | 112212146 |
| 22 | 301100754 |
| 23 | 805570061 |
| 24 | 2158326727 |
| 25 | 5768299665 |
| 26 | 9378272603 |
| 27 | $1.298824552 * 10^{10}$ |
| 28 | $1.659821848 * 10^{10}$ |
| 29 | $2.020819142 * 10^{10}$ |
| 30 | $2.381816436 * 10^{10}$ |
| 31 | $2.742813729 * 10^{10}$ |
| 32 | $3.194703212 * 10^{10}$ |
| 33 | $3.582466094 * 10^{10}$ |
| 34 | $3.646592696 * 10^{10}$ |
| 35 | $4.209591389 * 10^{10}$ |
| 36 | $5.224479567 * 10^{10}$ |
| 37 | $2.26774391 * 10^{11}$ |

| 38 | $2.502910876 \times 10^{11}$ |
|----|------------------------------|
| 39 | $3.600356717 \times 10^{11}$ |
| 40 | $4.305857616 \times 10^{11}$ |
| 41 | $1.026342076 \times 10^{12}$ |
| 42 | $3.636695399 \times 10^{12}$ |
| 43 | $1.026659714 \times 10^{30}$ |
| 44 | $9.810623334 \times 10^{44}$ |
| 45 | $7.823443122 \times 10^{48}$ |
| 46 | $3.026769726 \times 10^{55}$ |
| 47 | $4.568680244 \times 10^{60}$ |
| 48 | $8.726623484 \times 10^{66}$ |
| 49 | $2.246669926 \times 10^{70}$ |
| 50 | $3.325586728 \times 10^{74}$ |
| 51 | $9.568832546 \times 10^{77}$ |

*Table 2* -  Total number of conformations in simple cubic lattices for various chain lengths (N).

| Number of Amino acid residues (N) | Total Number of conformations |
|---|---|
| 2 | 25 |
| 3 | 121 |
| 4 | 589 |
| 5 | 2821 |
| 6 | 13565 |
| 7 | 64661 |
| 8 | 308981 |
| 9 | 1468313 |
| 10 | 6989025 |
| 11 | 33140457 |
| 12 | 157329085 |
| 13 | 744818613 |
| 14 | 3529191009 |
| 15 | 3529191009 |
| 16 | 16686979329 |
| 17 | 78955042017 |
| 18 | 372953947349 |
| 19 | 1762672203269 |
| 20 | 8319554639789 |
| 21 | 39285015083693 |
| 22 | 185296997240401 |
| 23 | 874331369198569 |
| 24 | 4121696814263465 |
| 25 | 19436473616738893 |
| 26 | 91582299477850157 |
| 27 | 431645810810533429 |
| 28 | 2033030798029977817 |
| 29 | 9577818849158639505 |
| 30 | 45094984254242445769 |
| 31 | 212365177454402817661 |
| 32 | 999559910134936147749 |
| 33 | 4705624045427280858353 |
| 34 | 22142271604470539035097 |
| 35 | 104208021575426329129665 |
| 36 | 490228476055783621093445 |
| 37 | 2305900954923345233920000 |
| 38 | 10845142436567878766570033 |
| 39 | 51001165482563218900000002 |
| 40 | 23981659235433476636670000 |

*Table 3* -Total number of designing sequences, designable conformations and non-designable sequences for various amino acid chain lengths (N) for square lattices. The chain length varies from 4 to 36. The designing sequences, non-designable sequences and designable conformations are calculated when p=N/2 and p= (N-1)/2 for even and odd values of N respectively.

| Number of Amino acid residues(N) | Number of conformations | Number of designing sequences($S_N$) | Number of designable conformations($D_N$) | Number of non-designable sequences |
|---|---|---|---|---|
| 4 | 5 | 4 | 1 | 8 |
| 5 | 13 | 0 | 0 | 20 |
| 6 | 36 | 7 | 3 | 29 |
| 7 | 98 | 10 | 2 | 39 |
| 8 | 272 | 7 | 5 | 57 |
| 9 | 740 | 6 | 4 | 73 |
| 10 | 2034 | 6 | 4 | 94 |
| 11 | 5513 | 62 | 14 | 110 |
| 12 | 15037 | 87 | 25 | 132 |
| 13 | 40617 | 173 | 52 | 156 |
| 14 | 110188 | 386 | 130 | 182 |
| 15 | 296806 | 857 | 218 | 210 |
| 16 | 802075 | 1539 | 456 | 241 |
| 17 | 2155667 | 3404 | 787 | 273 |
| 18 | 5808335 | 6349 | 1475 | 299 |
| 19 | 15582342 | 13454 | 2726 | 341 |
| 20 | 41889578 | 24900 | 5310 | 379 |
| 21 | 112212146 | 52183 | 9156 | 419 |
| 22 | 301100754 | 97478 | 17881 | 471 |
| 23 | 805570061 | 199290 | 31466 | 506 |
| 24 | 2158326727 | 380382 | 61086 | 548 |
| 25 | 5768299665 | 765147 | 107336 | 577 |
| 26 | 9378272603 | 1544312 | 219300 | 705 |
| 27 | $1.298824552 * 10^{10}$ | 3365700 | 391217 | 802 |
| 28 | $1.659821848 * 10^{10}$ | 6739200 | 758597 | 922 |
| 29 | $2.020819142 * 10^{10}$ | 13858416 | 1307420 | 10810 |
| 30 | $2.381816436 * 10^{10}$ | 27793600 | 2405004 | 11282 |
| 31 | $2.742813729 * 10^{10}$ | 53729200 | 4660252 | 13341 |
| 32 | $3.194703212 * 10^{10}$ | 109724224 | 9301076 | 14874 |
| 33 | $3.582466094 * 10^{10}$ | 278586544 | 18853380 | 16453 |
| 34 | $3.646592696 * 10^{10}$ | 444393600 | 36945640 | 18744 |
| 35 | $4.209591389 * 10^{10}$ | 877203600 | 72564464 | 19452 |
| 36 | $5.224479567 * 10^{10}$ | 1753614912 | 145675380 | 21655 |

*Table 4* -Total number of *most designable, less designable and designing sequences*, along with the number of *designable conformations* for various amino acid chain lengths (N) in square lattices. These numbers are valid when p=N/2 and p= (N-1)/2 for even and odd values of N respectively.

| Number of Amino acid residues(N) | Number of most designable sequences | Number of less designable sequences | Number of designing sequences($S_N$) | Number of designable conformations($D_N$) |
|---|---|---|---|---|
| 4 | 2 | 2 | 4 | 1 |
| 5 | 0 | 0 | 0 | 0 |
| 6 | 1 | 6 | 7 | 3 |
| 7 | 3 | 7 | 10 | 2 |
| 8 | 0 | 7 | 7 | 5 |
| 9 | 1 | 5 | 6 | 4 |
| 10 | 4 | 2 | 6 | 4 |
| 11 | 44 | 18 | 62 | 14 |
| 12 | 71 | 16 | 87 | 25 |
| 13 | 95 | 78 | 173 | 52 |
| 14 | 190 | 196 | 386 | 130 |
| 15 | 450 | 407 | 857 | 218 |
| 16 | 1027 | 512 | 1539 | 456 |
| 17 | 1381 | 2023 | 3404 | 787 |
| 18 | 3109 | 3240 | 6349 | 1475 |
| 19 | 4429 | 9025 | 13454 | 2726 |
| 20 | 8900 | 16000 | 24900 | 5310 |
| 21 | 34543 | 17640 | 52183 | 9156 |
| 22 | 74246 | 23232 | 97478 | 17881 |
| 23 | 162260 | 37030 | 199290 | 31466 |
| 24 | 131550 | 248832 | 380382 | 61086 |
| 25 | 236973 | 528174 | 765147 | 107336 |
| 26 | 369096 | 1175216 | 1544312 | 219300 |
| 27 | 1771377 | 1594323 | 3365700 | 391217 |
| 28 | 2260992 | 4478208 | 6739200 | 758597 |
| 29 | 7883111 | 5975305 | 13858416 | 1307420 |
| 30 | 11431600 | 16362000 | 27793600 | 2405004 |
| 31 | 20106657 | 33622543 | 53729200 | 4660252 |
| 32 | 54598272 | 55125952 | 109724224 | 9301076 |
| 33 | 178265230 | 100321314 | 278586544 | 18853380 |
| 34 | 219889152 | 224504448 | 444393600 | 36945640 |

| 35 | 340421200 | 536782400 | 877203600 | 72564464 |
| 36 | 1632682560 | 120932352 | 1753614912 | 145675380 |

*Table 5* - Total number of designing, non-designable sequences as well as designable conformations for various amino acid chain lengths (N) for simple cubic lattices. The chain length varies from 4 to 27. These values are valid when p=N/2 and p= (N-1)/2 for even and odd values of N respectively.

| Number of Amino acid residues (N) | Number of conformations | Number of designing sequences($S_N$) | Number of designable conformations($D_N$) | Number of non-designable sequences |
|---|---|---|---|---|
| 4 | 121 | 3 | 1 | 0 |
| 5 | 589 | 0 | 0 | 0 |
| 6 | 2821 | 0 | 0 | 0 |
| 7 | 13565 | 0 | 0 | 0 |
| 8 | 64661 | 2 | 2 | 3 |
| 9 | 308981 | 0 | 0 | 0 |
| 10 | 1468313 | 0 | 0 | 0 |
| 11 | 6989025 | 0 | 0 | 0 |
| 12 | 33140457 | 2 | 2 | 1 |
| 13 | 157329085 | 0 | 0 | 0 |
| 14 | 744818613 | 1 | 1 | 1 |
| 15 | 3529191009 | 1 | 1 | 0 |
| 16 | 16686979329 | 1 | 1 | 0 |
| 17 | 78955042017 | 8 | 8 | 4 |
| 18 | 372953947349 | 29 | 28 | 13 |
| 19 | 1762672203269 | 47 | 42 | 19 |
| 20 | 8319554639789 | 64 | 60 | 33 |
| 21 | 39285015083693 | 73 | 72 | 37 |
| 22 | 185296997240401 | 88 | 85 | 43 |
| 23 | 874331369198569 | 103 | 98 | 79 |

| 24 | 4121696814263465 | 121 | 104 | 91 |
|----|------------------|-----|-----|-----|
| 25 | 19436473616738893 | 133 | 124 | 103 |
| 26 | 91582299477850157 | 139 | 131 | 114 |
| 27 | 431645810810533429 | 157 | 149 | 121 |

*Table 6:* Total number of *most designable, less designable and designing sequences as well as designable conformations* for various amino acid chain lengths (N) in simple cubic lattices. These values are valid when p=N/2 and p= (N-1)/2 for even and odd values of N respectively.

| Number of Amino acid residues(N) | Number of most designable sequences | Number of less designable sequences | Number of designing sequences($S_N$) | Number of designable conformations($D_N$) |
|---|---|---|---|---|
| 4 | 2 | 1 | 3 | 1 |
| 5 | 0 | 0 | 0 | 0 |
| 6 | 0 | 0 | 0 | 0 |
| 7 | 0 | 0 | 0 | 0 |
| 8 | 2 | 0 | 2 | 2 |
| 9 | 0 | 0 | 0 | 0 |
| 10 | 0 | 0 | 0 | 0 |
| 11 | 0 | 0 | 0 | 0 |
| 12 | 2 | 0 | 2 | 2 |
| 13 | 0 | 0 | 0 | 0 |
| 14 | 1 | 0 | 1 | 1 |
| 15 | 1 | 0 | 1 | 1 |
| 16 | 1 | 0 | 1 | 1 |
| 17 | 4 | 4 | 8 | 8 |
| 18 | 16 | 13 | 29 | 28 |
| 19 | 24 | 23 | 47 | 42 |
| 20 | 20 | 44 | 64 | 60 |
| 21 | 16 | 57 | 73 | 72 |
| 22 | 45 | 43 | 88 | 85 |
| 23 | 44 | 59 | 103 | 98 |
| 24 | 56 | 65 | 121 | 104 |
| 25 | 63 | 70 | 133 | 124 |
| 26 | 66 | 73 | 139 | 131 |
| 27 | 60 | 97 | 157 | 149 |
| **64** | **108** | **281** | **389** | **156** |